\documentclass[11pt]{amsart}

\usepackage{amsfonts}
\usepackage{amssymb}
\usepackage{bm}

\newtheorem{thm}{Theorem}
\newtheorem{defn}[thm]{Definition}
\newtheorem{lemma}[thm]{Lemma}
\newtheorem{coro}[thm]{Corollary}

\def\R{\mathbb{R}}
\def\Z{\mathbb{Z}}
\def\C{\mathbb{C}}

\def\Hop{{\bf H}}

\let\imbeds=\hookrightarrow
\def\antipar{\mathbin{\upharpoonleft\!\!\!\;\downharpoonright}}
\def\dirpar{\mathbin{\upharpoonleft\!\!\!\;\upharpoonright}}

\def\abs#1{\mathopen|#1\mathclose|}
\def\Abs#1{\left|#1\right|}

\let\Dst=\displaystyle
\let\Tst=\textstyle

\mathcode`\`="8000
\begingroup
\makeatletter
\gdef\rqch@r{\hbox{\char'140}}
\catcode`\`=13
\gdef`{\@ifnextchar[{\mathclose}{\@ifnextchar]{\mathopen}{\@ifnextchar<{\mathopen}{\@ifnextchar>{\mathclose}{\rqch@r}}}}}
\makeatother
\endgroup

\def\avint{
\mathchoice{\hbox to 0pt{\,\tenln\char8\hss}}
           {\hbox to 0pt{\raise-4pt\hbox to 1.8pt{\char19\hss}%
                         \raise -2.6pt\hbox{\char19\hss}\hss}}%
           {\hbox to 0pt{\raise-4pt\hbox to 1.8pt{\char19\hss}%
                         \raise -2.6pt\hbox{\char19\hss}\hss}}%
           {\hbox to 0pt{\raise-4pt\hbox to 1.8pt{\char19\hss}%
                         \raise -2.6pt\hbox{\char19\hss}\hss}}%
\int}

\let\origmarginpar=\marginpar
\def\marginpar#1{\origmarginpar{\raggedright\footnotesize#1}}
\let\boundary=\partial

\def\qed{\hfill$\square$}

\def\n{\bm{n}}
\def\x{\mbox{\boldmath$x$}} 
\def\x{\bm{x}} 
\def\y{\bm{y}} 

\def\t{\mbox{\boldmath$t$}} 
\let\k=\kappa               
\def\psiv{\mbox{\boldmath$\psi$}} 
\def\psiv{\bm{\psi}} 
\def\muv{\mbox{\boldmath$\mu$}} 

\let\eps=\varepsilon

\def\RQ{R\!Q}
\def\sgn{\mathop{\rm sgn}}

\def\argmax{\mathop{\rm arg\,max}}

\def\BurTh{MR2203162}
\def\Linde{MR2240676}
\def\BenLo{MR2091490}
\def\Rocka{MR0274683}

\begin{document}

\author{Jochen Denzler}
\title{Existence and Regularity for a Curvature Dependent Variational Problem}
\email{denzler@math.utk.edu}
\address{Jochen Denzler, Math Dept, University of Tennessee,
  Knoxville, TN 37996, USA}


\thanks{%
The author gratefully acknowledges repeated useful discussions with
Almut Burchard. She inspired the research and was available to discuss
and critique progress. 
Some ideas of hers enter in the arguments, as outlined in the main
text. Reseach was partly supported by a grant from the Simons
foundation (\#208550). The hospitality of the CRM Universit\'e de
Montr\'eal during the workshop on Geometry of Eigenfunctions June 4-8,
2012 was a boost to this research, as was a Faculty Development Leave
(`Sabbatical') the University of Tennessee granted me during Spring
2012, and the hospitality of Karlsruhe Institute of Technology during
said leave.  
}

\begin{abstract}
 It is proved that smooth closed curves of given length minimizing the
 principal eigenvalue of the Schr\"odinger operator
 $-\frac{d^2}{ds^2}+\kappa^2$ exist. Here $s$ denotes the arclength and
 $\kappa$ the curvature. These minimizers are automatically
 planar, analytic, convex curves. The straight segment, traversed back
 and forth, is the only possible exception that  becomes admissible in a
 more generalized setting. In proving this, we overcome the difficulty
 from a lack of coercivity and compactness by a combination of 
 methods. 
\end{abstract}

\maketitle

\section{Introduction and Outline}

Given a closed curve $\gamma:s\mapsto \x(s)$ of length $2\pi$
in~$\R^n$, we consider  
the principal eigenvalue $\lambda$ of the Schr\"odinger operator
$\Hop= -\frac{d^2}{ds^2}+\kappa^2$ on the curve; here $s$ is the
arclength and $\kappa(s)$ is the curvature
$\abs{\x''(s)}$. Considering $\lambda$ as a function of the curve, we
ask what (if any) is the minimal possible value of $\lambda$, and for
which curves it is taken on. 
Fixing the length to $2\pi$ is no loss of generality, since the
problem is homogeneous with respect to dilations.

A natural conjecture is that the minimal $\lambda$ is~1; this value is
indeed the principal eigenvalue of~$\Hop$ for the unit circle. But it
is known to be also the principal eigenvalue of $\Hop$ for a certain
1-parameter family of ovals (convex planar curves); this family
connects the unit circle to a limiting case we call `di-gon': a
straight segment of length $\pi$ traversed back and forth. 

The Oval Conjecture states that this family of ovals does consist of
minimizers. A strengthened version would stipulate that these are the
only minimizers. So far, the Oval Conjecture is open. 
It has been shown by Burchard and Thomas \cite{\BurTh} that this family
consists of {\em relative\/} minimizers (in the sense of weak
minimizers: minimal among competitors in a neighborhood defined by a
strong topology). It is also easy to see (from the equivalent version
(\ref{relOPharm}) below and the fact that each component $\psi_i$ there
must have a zero) that 
$\lambda\ge\frac14$ in any case, and that $\lambda\ge1$ for curves
enjoying a point symmetry. An elegant elementary argument (for convex
planar curves) by Linde \cite{\Linde} raised the
previously known lower bound $\lambda\ge\frac12$ \cite{\BenLo}
to some quantity
$\lambda_*\approx0.6085$, and also established $\lambda\ge1$ for a
certain class of ovals defined in geometric terms.

An interesting aspect of this variational problem is that a positive
answer to the above conjecture (in 2 dimensions) implies that the best
constant $L$ in the 1-dimensional Lieb-Thirring inequality $\lambda\le L
\int_{\R} V_-^{3/2}$ for the Schr\"odinger operator
$-\frac{d^2}{dx^2}+V(x)$ with a {\em single\/} bound state also
applies to potentials with two bound states. See \cite{\BenLo} for
details. (Lest a wrong impression be created by omission, it may be
mentioned here that the connection in \cite{\BenLo} does {\em not\/}
identify $V$ with $\kappa^2$). 

Moreover, Bernstein and Breiner \cite{BernBr} have established a
connection between the Oval Problem and a minimization property of the
catenoid. Namely, they show that among all {\em minimal\/} surfaces of
the topological type of an annulus that connect two parallel planes in
$\R^3$, the marginally stable catenoid has the smallest area. One of
the proofs they provide relies on the assumption that the oval
conjecture holds; however they also give a proof that is independent
of the oval conjecture. 

The degenerating family of ovals with common principal eigenvalue~1
shows a lack of compactness (even of coercivity) in the
problem. Sublevel sets $\{\gamma\mid\lambda(\gamma)\le a\}$ for
$a\ge1$ lack any a-priori bounds on curvature, even in an $L^p$ norm
with $p>1$. Therefore they fail to be compact in any plausible
topology for the problem. This is an obstruction to an existence proof
by direct methods, and is also an obstruction to a regularity theory
for minimizers, should they indeed exist. 

Despite these difficulties, we prove in this paper existence and
regularity for minimizers. 
Incidentially, this shows that Linde's assumptions in his proof that
$\lambda\ge \lambda_*\approx0.6085$ for planar convex curves 
are no loss of generality in the full problem.

Specifically we prove:
\begin{thm}
Among all closed $W^{2,2}$ (or $C^2$) curves of length $2\pi$ in
$\R^n$, there exists one that minimizes the principal eigenvalue of
the operator $\Hop=-\frac{d^2}{ds^2}+\kappa^2$, where $s$ is the
arclength and $\kappa$ is the curvature. Minimizing curves are planar
convex analytic curves with strictly positive curvature. 
\end{thm}

To achieve this goal, we first define a relaxed variation problem for
which some compactness is restored, allowing for an existence
proof. We then show, comparatively easily, that the singular set of a
potential minimizer can consist of at most 2 points. An explicit
descent algorithm shows that any minimizer must be planar and
convex. The Euler-Lagrange equations of the variation problem play
only a weak role in this argument: we basically exploit them for
standard regularity results for solutions to ODEs. Subsequently, we
will also use them for local asymptotic results near singularities. 
Beyond this use, the Euler-Lagrange equations so far seem to be of very
limited use in this problem. Indeed, they appear to allow for
chaotic dynamics (loosely speaking, and judging merely on the basis of
some numerical experiments). 
In particular, planarity is {\em not\/} a feature that
would follow from the Euler-Lagrange dynamics. However, Euler-Lagrange
dynamics does imply that extremal curves lie in a space of dimension at
most~3. While this fact is not essential for our argument, it does
simplify the reasoning a bit. 

The arguments outlined so far still amount only to a partial
regularity result, leaving the possibility of `D-shaped'
minimizers. We call a minimizing curve D-shaped if it has one or two
singular points, where the curvature fails to be $C^2$ (possibly even
allowing for a corner there), and where a straight segment connects
the two singular points; the case of only one singular point is
included, with the straight segment then omitted.

We will rule out D-shaped minimizers by a combination of two tools,
namely: 
Asymptotic analysis near the singularity, following from the
Euler-Lagrange equation (albeit with some a-priori input
derived from minimality); and some further curve surgery argument that
applies to minimizers only. These latter surgery arguments are
of a local nature (i.e., variations supported on small intervals) and
amount to strong variations not seen by the Euler-Lagrange
equations. In this context, strong variations are those that are small
in the $C^1$ norm, but large in the $C^2$ norm of the curve. 

In the relaxed variation problem, the di-gon is a possible
minimizer (that is ruled out by the $W^{2,2}$ assumption on the
curve). Should the minimum of the principal eigenvalue be 1 as
conjectured, then the above-mentioned explicitly known ovals that have
the di-gon as a limiting case are
regular minimizers (we do not know if they are the only ones). Should
the minimum be less than~1, the sole possible exception to regularity
for the relaxed variation problem doesn't apply at all. 

For the detailed proof, the above theorem is split up into
Thms.~\ref{Existence}, \ref{planar}, and \ref{FullRegularity} below,
which are proved separately.

While the oval conjecture remains open, certainly the existence
theorem and geometric properties of minimizers proved here should
be expected to limit the quest for minimizers in a useful manner.

\section{Definition of Original and Relaxed Variation Problems}

We write $S^1=\R/2\pi\Z$ for the unit circle  and assume we have a
closed rectifiable curve $\gamma$ parametrized by arclength~$s$, and
with reasonably defined curvature.  So $\gamma$ is 
given by $\x\in C^2(S^1\to\R^n)$, or possibly $\x\in
W^{2,2}(S^1\to\R^n)$, subject to $\abs{\x''}\equiv1$. We let
$\kappa(s):= \abs{\x''(s)}$ denote the curvature. The principal
eigenvalue $\lambda$ of the operator
$\Hop=-\frac{d^2}{ds^2}+\kappa^2(s)$   is defined by the Rayleigh-Ritz
variation problem
\begin{equation}\label{RR} 
\lambda(\gamma) := \min \left\{
\int_0^{2\pi} \!\bigl(\phi'^2(s)+\kappa^2(s)\phi^2(s)\bigr)\,ds \Bigm| 
\phi\in W^{1,2}(S^1) \,,\;\, \int_0^{2\pi}\phi^2(s)\,ds=1 \right\}
\,,
\end{equation}
which is known to have a minimizer that is unique up to sign; we
can choose $\phi\ge0$ with no loss of generality. Then, with
$\kappa^2\in L^1$, the eigenfunction  $\phi$ lies in
$W^{2,1}\subset C^1$ and satisfies
$-\phi''+\kappa^2\phi=\lambda\phi$. Nonnegative solutions of this
equation cannot vanish at all unless they vanish identically; this
follows from a version of the uniqueness theorem for initial value
problems for linear ODEs with $L^1$ coefficients. 
We are studying the minimization problem
\begin{equation}\label{OP}
\inf\left\{ \lambda(\gamma) \Bigm|
\gamma: s\mapsto \x(s) \mbox{ with } \x\in
W^{2,2}(S^1\to\R^n)\,,\; \abs{\x'}\equiv1\,,\; \abs{\x''(s)}=\kappa(s)
\right\}
\;.
\end{equation}
These two can be combined into the variation problem
\begin{equation}\label{ClassOPcurve}
\begin{array}{l}
I[\x,\phi]:= \int_0^{2\pi} \bigl(\phi'^2+\abs{\x''}^2\phi^2\bigr)\,ds  
\,,
\\[1.5ex]\Dst
\inf\left\{ I[\x,\phi]
\bigm|
\x\in W^{2,2}(S^1\to\R^n)\,,\; \phi\in W^{1,2}(S^1)\,,\;
\abs{\x'}\equiv1\,,\; \|\phi\|_{L^2}^2=1 
\right\}
\;.
\end{array}
\end{equation}
As outlined, it is no loss of generality also to require $\phi>0$
in~(\ref{ClassOPcurve}). 
Introducing the function $\psiv:=\x'\phi\in W^{1,2}(S^1\to\R^n)$, there
is an equivalent formulation that was
already used in \cite{\BurTh}. 
The condition that $\x$ represents a closed curve requires
$\x'=\psiv/\abs{\psiv}$ to integrate to 0. 

\begin{defn}
The classical oval problem in curve coordinates is the variation
problem (\ref{ClassOPcurve}). The classical oval problem in harmonic
coordinates is
\begin{equation}\label{ClassOPharm}
\begin{array}{l}
I[\psiv]:= \int_0^{2\pi} \abs{\psiv'}^2(s)\,ds
\,,
\\[1.5ex]
\inf\left\{  I[\psiv]
\bigm|
\psiv\in
W^{1,2}(S^1\to\R^n\setminus\{{\bf0}\})\,,\; \int_0^{2\pi}\abs{\psiv}^2\,ds=1 \,,\; 
\int_0^{2\pi}\frac{\psiv}{\abs{\psiv}}\,ds={\bf0}
\right\}
\;.
\end{array}
\end{equation}
\end{defn}

The following establishes the equivalence of (\ref{ClassOPcurve}) and
(\ref{ClassOPharm}) and is routine to check: 
\begin{lemma}\label{LemEQclass}
If $(\x,\phi)$ is in the domain of (\ref{ClassOPcurve}) and $\phi>0$,
then $\psiv:=\x'\phi$ is in the domain of
(\ref{ClassOPharm}). Conversely, if $\psiv$ is in the domain of
(\ref{ClassOPharm}), then $\phi:=\abs{\psiv}$
and $\x:=\int\psiv/\abs{\psiv}$ (with any choice of the constant of
integration) provide an $(\x,\phi)$ in the domain of
(\ref{ClassOPcurve}). It holds: 
$I[\x,\phi]=I[\psiv]$. 
\end{lemma}

Now in (\ref{ClassOPcurve}), the functional does not control
$\int\abs{\x''}^2$. In (\ref{ClassOPharm}), the condition that
$\psiv$ doesn't vanish anywhere is not stable under any sensible convergence
notion for $\psiv$, in particular not weak or strong $W^{1,2}$
convergence. It does make sense to enlarge the domain of
(\ref{ClassOPharm}) by requiring $\psiv$ to vanish only on a set of
measure~zero, but this does not mend the loss of stability under
convergence. Doing so allows curves $\x$ with non-$L^2$
curvature, or even with corners, provided $\phi$ vanishes in those
points. 

Given $\x\in W^{1,\infty}(S^1)$, we define with the obvious
understanding of an interval $[s-\eps,s+\eps]$ as a subset of $S^1$,
the {\em exceptional set\/} 
$$
E[\x]:= \{s \mid \x\notin W^{2,2}[s-\eps,s+\eps] \mbox{
  for any $\eps$ }\}
\;.
$$
Given $\psiv\in W^{1,2}(S^1\to\R^n)$, we define the zero set
$$
Z[\psiv]:= \{s \mid \psiv(s)={\bf0}\}
\;.
$$

By definition of $E$ and continuity of $\psiv$ respectively, $E[\x]$
and $Z[\psiv]$ are closed sets; $\k$ is defined exactly on the 
complement of~$E[\x]$.

\begin{defn}
The relaxed oval problem in curve coordinates is
\begin{equation}\label{relOPcurve}
\begin{array}{l}
I[\x,\phi]:= 
\int_{S^1\setminus E[\x]} \bigl(\phi'^2+\abs{\x''}^2\phi^2\bigr)\,ds 
\,\mbox{  for  } \,\x\in W^{1,\infty}(S^1\to\R^n)\,,\,\, 
\phi\in W^{1,2}(S^1)
\\[1.5ex]
\inf\left\{ I[\x,\phi]
\bigm|
\abs{\x'}\equiv1\,,\; \|\phi\|_{L^2}^2=1 \,,\; \phi=0\mbox{ on }E[\x]
\right\}
\;.
\end{array}
\end{equation}
We let $\sigma$ denote the 1-dimensional Lebesgue measure and define
$\sgn\psiv:= \psiv/\abs{\psiv}$ provided $\psiv\neq{\bf0}$, and
$\sgn{\bf0}:={\bf0}$.  
Then the relaxed oval problem in harmonic coordinates is 
\begin{equation}\label{relOPharm}
\begin{array}{l}
I[\psiv]:= \int_0^{2\pi} \abs{\psiv'}^2(s)\,ds 
\\[1.5ex]
\inf\left\{ I[\psiv]
\bigm|
\psiv\in
W^{1,2}(S^1\to\R^n)\,,\; \int_0^{2\pi}\abs{\psiv}^2\,ds=1 \,,\; 
\Abs{
\int_0^{2\pi}\sgn\psiv\,ds} \le \sigma(Z[\psiv])
\right\}
\end{array}
\end{equation}
We refer to this last constraint as the weak loop condition. 
\end{defn}

Note that the given domain in (\ref{relOPcurve}) ascertains that the
integral $I[\x,\phi]$ in the functional is defined in the extended sense, but 
does not guarantee a finite value for it.

We now claim
\begin{lemma}\label{LemEQrel}
If $(\x,\phi)$ is in the domain of (\ref{relOPcurve}) with
$I[\x,\phi]<\infty$,  then $\psiv:=\x'\phi$ is in the domain of
(\ref{relOPharm}), $Z[\psiv]\supset E[\x]$,
and the functional is the same, and $I[\psiv]=I[\x,\phi]$. 

Conversely, assume that $\psiv$ is in the domain of
(\ref{relOPharm}), let $\phi:=\abs{\psiv}$ 
and $\y:=\int\sgn{\psiv}$ (with any choice of the constant of
integration). Then $\y$ describes a (not necessarily  closed)
rectifiable curve segment of length $\sigma(Z[\psiv]^c)$ that can be
extended to a closed curve~$\x$ of length $2\pi$, such
that $E[\x]\subset Z[\psiv]$ and $(\x,\phi)$ is in the domain of
(\ref{ClassOPcurve}), and $I[\psiv]=I[\x,\phi]$. 
\end{lemma}

{\sc Proof: }
If $(\x,\phi)$ is in the domain of (\ref{relOPcurve}), then $E[\x]^c$
cannot be empty because this would entail incompatible constraints
on~$\phi$. By definition of $E[\x]$, we have $\x\in
W^{2,2}_{\rm{loc}}(E[\x]^c)$, hence $\psiv=\x'\phi\in
W^{1,2}_{\rm{loc}}(E[\x]^c)$. Since
$\abs{\psiv'}^2=\abs{\phi'}^2+\abs{\x''}^2\phi^2$, the finiteness of
the functional in~(\ref{relOPcurve}) implies $\psiv\in
W^{1,2}(E[\x]^c)$. So $\psiv$ is continuous on $E[\x]^c$. 
Moreover $\abs{\psiv}=\abs{\phi}$ is continuous on~$S^1$ and
vanishes on $E[\x]$. So $\psiv\in C^0(S^1)$ and
$\psiv={\bf0}$ on $E[\x]$. The extension of $\psiv\in
W^{1,2}(E[\x]^c)$ by {\bf0} to $S^1$ is therefore in
$W^{1,2}(S^1)$. We have also obtained $Z[\psiv]\supset E[\x]$ in the
process and only have to verify the weak loop condition yet. 
But $\abs{\int_{S^1}\sgn\psiv} = \abs{\int_{Z[\psiv]^c}\x'}
= \abs{-\int_{Z[\psiv]}\x'}\le\sigma(Z[\psiv])$. 

For the converse statement,
assume $\psiv$ is in the domain of (\ref{relOPharm}). Clearly
$\psi:=\abs{\psiv}$ has $L^2$-norm~1, is continuous on
$S^1$ and vanishes on $Z[\psiv]$, because $\psiv\in W^{1,2}(S^1)$ has
these properties. Also $\psi\in W^{1,2}(Z[\psiv]^c)$. Hence $\psi \in
W^{1,2}(S^1)$ with $\psi,\psi'=0$ on $Z[\psiv]$. The function $\phi$
will shortly arise from $\psi$ by extension and reparametrization.  

Next, $\y(t):= \int_0^t\sgn\psiv(t)\,dt$ defines a 
function in $W^{1,\infty}([0,2\pi]\to\R^n)$ with
$\abs{\frac{d}{dt}\y}=1$ on $Z[\psiv]^c$ and $\abs{\frac{d}{dt}\y}=0$
on $Z[\psiv]$. This function~$\y$ 
represents a curve segment of length exactly $\sigma(Z[\psiv]^c)$, with the
arc length parameter not $t$, but $s:=\int_0^t\abs{\sgn\psiv(t)}\,dt$. 
We estimate $\abs{\y(2\pi)-\y(0)}
= \abs{\int_0^{2\pi} \y'(t)\,dt}
=\abs{\int_0^{2\pi}\sgn\psiv(t)\,dt} \le \sigma(Z[\psiv])$, where the weak
loop condition from (\ref{relOPharm}) was used in the last step. 
So the curve segment $\y$ can be extended to a closed curve $\gamma$  by adding
a smooth curve of length {\em exactly\/} $\sigma(Z[\psiv])=:\ell$ with
$t\in[2\pi,2\pi+\ell]$ the arclength parameter on this segment. 

We extend $\psi$ by 0 on this extra piece of curve. So 
$\y$ and $\psi$ are now functions of $t\in[0,2\pi+\ell]$. 
Reparametrizing them to arclength $s\in[0,2\pi]$ according to 
$s=\int_0^t\abs{\sgn\psiv(t)}\,dt$ for $t\in[0,2\pi]$, and
$s=t-\ell$ for $t\in[2\pi,2\pi+\ell]$, 
we get $\x(s):=y(\t)$, $\phi(s):=\psi(t)$. 

We have $\int(\frac{d\phi}{ds})^2\,ds = \int (\frac{d\psi}{dt})^2\,dt$
and $\int \phi(s)^2\,ds = \int \psi(t)^2\,dt$, since
$\frac{ds}{dt}=1$ wherever $\phi\neq0$. Since $\x\in W^{2,2}_{\rm
loc}(Z[\psiv]^c)$, we know $E[\x]\subset Z[\psiv]$, as far as
$s\in[0,\sigma(Z[\psiv]^c)$ is concerned. Clearly
$s\in`]\sigma(Z[\psiv]^c),2\pi`[$ does not contain any points of
$E[\x]$, since we chose a smooth connecting segment. 

Equality of the functionals applies for the same reason as in the
comparison of~$\psi$ with~$\phi$.
\qed

We will often switch between harmonic and curve coordinates, based on
Lemmas~\ref{LemEQclass} and \ref{LemEQrel}, without further
comment. We may also abandon the normalization conditions
$\|\phi\|_{L^2}=1$, $\|\psiv\|_{L^2}=1$ 
and minimize the Rayleigh quotients $I[\x,\phi]/\|\phi\|_{L^2}^2$, 
$I[\psiv]/\|\psiv\|_{L^2}^2$ respectively. 
We may also abandon the length constraint and minimize
$(\mbox{length}/2\pi)^2$ times the Rayleigh quotient instead. 

\section{Existence and Partial Regularity for the Relaxed Oval
Problem}

We are now ready to prove
\begin{thm}[Existence and Classification]\label{Existence}
The variational problem (\ref{relOPharm})
has a minimum~$\psiv$ with $I[\psiv]\le1$.  
The following alternative exists for the 
zero set $Z[\psiv]$ of such a minimizer: 

{\em Either (a) }
$Z[\psiv]=\emptyset$.  In this case, the associated curve has
everywhere defined curvature in the $L^2$ sense, i.e.,
$\x\in W^{2,2}$. 

{\em Or (b) }
$Z[\psiv]$ is a closed interval of length $<\pi$, possibly
degenerating to a single point. In this case the weak loop condition
is satisfied with equality, $\x$ contains a 
straight segment along $Z[\psiv]$ (omitted if $Z[\psiv]$ is a
singleton), with $\phi\equiv0$ on $Z[\psiv]$, and  
$\x$ has everywhere defined curvature in the
$W^{2,2}_{loc}(Z[\psiv]^c)$ sense.   

{\em Or (c1) }
$Z[\psiv]$ is a 
closed interval of length exactly $\pi$.  In this case, the associated
curve is a `di-gon' (i.e.\ a line segment traversed once back and forth),
with $\phi$ supported on one side of the digon. 

{\em Or (c2) }
$Z[\psiv]$  consists of two points with length exactly $\pi$ apart. In
this case, the associated curve is the same `di-gon', but $\phi$ is 
supported on both segments. 
\end{thm}
\begin{defn}
Minimizers in the case (b) of the preceding theorem will be called D-shaped.
\end{defn}

In the present section we will prove this theorem, and also show that
minimizers must be convex planar curves, more specifically:
\begin{thm}[Planarity and Convexity]\label{planar}
Minimizers $\psiv$ of (\ref{relOPharm}) with $Z[\psiv]=\emptyset$ 
represent planar, strictly convex, real-analytic curves with strictly
positive curvature.  

D-shaped minimizers $\psiv$ (if any) represent planar convex curves,
the portion over $Z[\psiv]^c$ of which is real-analytic and  
has strictly positive curvature.
\end{thm}

However, it is
worth announcing already now the stronger result proved in
Section~\ref{no-D-shaped} that case (b) of Thm.~\ref{Existence} does
not occur:  
\begin{thm}[Regularity]\label{FullRegularity}
With the possible exception of the di-gon, minimizers for the
variational problem (\ref{relOPharm}) are smooth, and $Z[\psiv]$ is empty. 
\end{thm}
The proof of this theorem will rely on the a-priori conclusions about
hypothetical D-shaped minimizers that are proved in the present
section. 

{\sc Proof of Thm.~\ref{Existence}: }
Assume $(\psiv_n)$ is a minimizing
sequence, i.e., $I[\psiv_n]\to\inf I$, and hence bounded in
$W^{1,2}$. We can extract a subsequence, again 
called $(\psiv_n)$, that converges uniformly, and weakly in $W^{1,2}$, to a
limit $\psiv_*$. We get
$I[\psiv_*]=\int|\psiv_*'|^2\le\liminf\int|\psiv_n'|^2=\inf I$ as well as
$\int|\psiv_*|^2=1$ routinely. We only need to show that $\psiv_*$
still satisfies the weak loop constraint.  
We find it convenient to tacitly adopt
the practice of using the notation $\sgn\psiv$ only in those cases
where the possibility of $\psiv$ 
vanishing need to be reckoned with, but to revert to $\psiv/|\psiv|$
in cases, where this possibility has already been ruled out. 
We also abbreviate $Z[\psiv_n],
Z[\psiv_*]$ as $Z_n,Z_*$ respectively.

Now clearly, on $Z_*^c$, it holds $\sgn\psiv_n\to\psiv_*/|\psiv_*|$ pointwise,
and trivially majorized. So we conclude 
\begin{equation}\label{sgnpsiL1conv}
  \int_{Z_*^c}\sgn\psiv_n \to
    \int_{Z_*^c}\psiv_*/|\psiv_*|
\;.
\end{equation}
On the other hand, 
$$
\int_{Z_*^c}\sgn\psiv_n 
= \int_{Z_*^c\cap Z_n^c} \psiv_n/|\psiv_n| 
= \int_{Z_n^c}\psiv_n/|\psiv_n| - \int_{Z_*\setminus Z_n} \psiv_n/|\psiv_n| 
$$
and therefore, using the fact that $\psiv_n$ satisfies the weak loop condition, 
$$
\Bigl|\int_{Z_*^c}\sgn\psiv_n\Bigr| \le
\Bigl|\int_{Z_n^c}\psiv_n/|\psiv_n|\Bigr| + \sigma(Z_*\setminus Z_n)
\le
\sigma(Z_n) +  \sigma(Z_*\setminus Z_n) = \sigma(Z_*)
\;.
$$
By taking the limit on the left, using~(\ref{sgnpsiL1conv}), we get
$\bigl|\int_{Z_*^c}\psiv_*/|\psiv_*|\big| \le\sigma(Z_*)$ as required.

The unit circle with constant eigenfunction, namely
$\psiv(s)= (2\pi)^{-1/2}\left[{\sin s\atop\cos s}\right]$ in $\R^2$
(or in $\R^n$, by imbedding) is an example with $I=1$, so clearly $\min
I\le1$. 

Having thus proved the existence of a minimizer, we now can get some limited
regularity. Let $\psiv$ be a minimizer and $Z$ its zero set. Its complement
$Z^c$, being open and non-empty, is either all of $S^1$ or is the
union of finitely many or 
countably infinitely many intervals $J_j$ with respective lengths $\ell_j$. 

With case (a) already being obvious from Lemma~\ref{LemEQrel}, 
let's look at the cases where  $Z\neq\emptyset$. 
Restricting $\psiv$ to $J_j$, this interval contributes
$\RQ_j:= \int_{J_j}|\psiv'|^2/\int_{J_j}|\psiv|^2$ to the Rayleigh
quotient
$\RQ=\int_0^{2\pi}\abs{\psiv'}^2/\int_0^{2\pi}\abs{\psiv}^2$. 
As~$\RQ$ is a weighted average of the local
Rayleigh quotients $\RQ_j$, it could be lowered by changing $\psiv$ to
{\bf0} on $J_j$, if 
$\RQ_j>\RQ$. So for a minimizer $\psiv$, all local $\RQ_j$ have to be equal,
namely $=\RQ[\psiv]\le1$. But because of the Dirichlet BC's
$\psiv={\bf0}$ on $\boundary I_j$, we know $\RQ_j\ge
(\pi/\ell_j)^2$, and so we  conclude $\ell_j\ge\pi$. So we have either 
a single $J_j$ of length $\ge\pi$, or two
$J_j$'s of length exactly $\pi$. Now when $\sigma(J_1)=\sigma(J_2)=\pi$, then  
only $\k\equiv0$
achieves a Rayleigh quotient 1, which is an upper bound for the
minimum. This leads to case (c2). 

If $Z^c$ consists of a single interval $J_1$ of length $\pi$, the restriction
$\x|_{J_1}$ must still be a straight segment by the same reasoning,
and the weak loop constraint forces $\x|_{Z}$ to be a straight
segment, too. So this is again the di-gon case in variant (c1).

Now let $Z^c$ consist of a single interval of length $\in`]\pi,2\pi]$, hence $Z$
is a closed interval of length $<\pi$, possibly degenerated to a
point. Then $\x$ consists of a $W^{2,2}_{\rm loc}$  
curve segment parametrized over the closure of 
$Z^c$, with $\phi$ supported there, and another segment closing the
curve, on which $\phi$ vanishes. If the weak loop constraint were
satisfied with slack, we could shorten the total length by replacing
the segment over $Z$ with a shorter straight segment, without
changing either $I[\x,\phi]$ or $\|\phi\|_{L^2}$. Then dilating the
curve $\x$ back to length $2\pi$, and rescaling $\phi$  to unit
$L^2$ norm, we would decrease $I[\x,\phi]$, contradicting the
minimality of the original curve. So the weak loop constraint must
have been satisfied with equality. By the strict triangle inequality,
the segment over $Z$ must then be straight.  
This is case (b). 
\qed

\medskip

Introducing a Lagrange multiplier $\lambda$ for the normalization
constraint  $\int\abs{\psiv'}^2=1$ and a vector valued Lagrange
multiplier $\muv$ for the loop constraint
$\int\psiv/\abs{\psiv}={\bf0}$, we routinely get that a minimizer
according to Case (a) of Thm.~\ref{Existence} must satisfy the Euler
Lagrange equation
\begin{equation}\label{EL}
\psiv'' +\lambda\psiv - (\muv\cdot\psiv) \psiv/|\psiv|^3
+ \muv/|\psiv| = 0
\,,
\end{equation}
or, in other words, 
$$
\psiv'' +\lambda\psiv +\frac{1}{|\psiv|} {\rm pr}_{\psiv\perp}\muv = 0
\,,
$$
where we have introduced the orthogonal projection of $\muv$ onto the
orthocomplement of $\psiv$. 
A~minimizer $\psiv_*$ according to Case (b) in Thm.~\ref{Existence} would in
particular minimize $I[\psiv]$ in the restricted class of those
$\psiv$ for which $Z[\psiv]=Z[\psiv_*]=:Z$ and
$\int_{Z^c}\psiv/\abs{\psiv}= \int_{Z^c}\psiv_*/\abs{\psiv_*}$. So the
same EL equations still hold in this case on the open interval 
$J:= Z^c$, together with the boundary conditions $\psiv={\bf0}$ on
$\boundary J$. 
Stronger information about $\psiv$ near $\boundary J$ will be
obtained below, eventually ruling out Case (b) altogether. 

Testing the equation (\ref{EL}) with $\psiv$ shows that the Lagrange
multiplier 
$\lambda$ for a minimizer $\psiv$ indeed coincides with the value
$\lambda=\min I$.  

As already observed in~\cite{\BurTh}, Equation (\ref{EL}) has a
conserved quantity  
\begin{equation}\label{Energy}
E= \frac12|\psiv'|^2 + \frac\lambda2|\psiv|^2 +\frac{\muv\cdot\psiv}{|\psiv|}
\;,
\end{equation}
and in higher dimensions than 2, there are lower dimensional angular momenta
arising from the rotation symmetry about the $\muv$ axis. 
The vector space spanned by $\psiv(s_0)$, $\psiv'(s_0)$ and $\muv$ is
invariant, so solutions of the EL equations automatically remain in an
at most 3-dimensional subspace of $\R^n$; possible connecting straight
segments
in the case of D-shaped minimizers would not leave this space either. 
Therefore minimizers are curves in at most a 3-dimensional space,
regardless of the dimension $n$ in which the Oval Problem was
originally posed. 
It also follows from standard regularity results for ODEs that
minimizers (or any extremals) are real-analytic curves on the
complement of the zero set $Z[\psiv]$. 

In the case of 2 dimensions, it is convenient to write the EL
equations and the energy equation in polar coordinates, and to
identify $\R^2$ with $\C$, where the real axis is chosen parallel to 
the vector $\muv$. 
It is therefore no loss of generality to choose
$\muv\in\R^2\cong\C$ to be a real nonnegative number $\mu$.
Then, writing $\psiv(s)=R(s)\exp i\theta(s)$, the EL equations and
energy become 
\begin{equation}\label{ELpolar}
\begin{array}{ll} 
R'' + R(\lambda-\theta'^2) = 0  & \quad \mbox{(note $\theta'=\k$ and $R=\phi$)}
\\
R\theta'' + 2R'\theta' = \frac{\mu}{R}\sin\theta  & \quad\mbox{(can also write
  as $(R^2\theta')'=\mu\sin\theta$)}
\\[2ex]
E= \frac12(R'^2+R^2\theta'^2+\lambda R^2) +\mu\cos\theta
\end{array}
\end{equation}
The first equation returns us the Schr\"odinger equation on the loop; the
second equation describes the interaction between Schr\"odinger eigenfunction
and curve that is necessary for an extremal.

The following simple geometric lemma will be useful in proving planarity of
minimizers. 
\begin{lemma}\label{L-curvetouch}
Given a closed $C^1$ curve $\gamma:s\mapsto\x(s)$ in $\R^n$ (not
necessarily injective), where $n\ge3$,  
there exists a hyperplane $\Pi$ that is tangential to $\gamma$ at
least twice, i.e., at points $\x(s_1)$, $\x(s_2)$ with $s_1\neq s_2$.

Given a closed $C^1$ curve $\gamma:s\mapsto\x(s)$ in $\R^2$ (not
necessarily injective), either there exists a line $\Pi$ that is
tangential to~$\gamma$ at least twice, or else, the tangent angle $\theta$
is a strictly monotonic function of $s$ with
$\theta(s+2\pi)=\theta(s)+2\pi$ or $\theta(s+2\pi)=\theta(s)-2\pi$,
with the sign depending on orientation. 
\end{lemma}
While we doubt that this lemma would be new, we do not have a 
reference for it and provide a proof; actually we give two different
proofs since both are interesting in their own right. 
Use of this lemma was inspired by Almut Burchard, and the proof by
convexity uses her ideas. We only need the lemma for $n\le3$. 

{\sc Proof of Lemma~\ref{L-curvetouch} (by convexity): }
Let $K$ be the convex hull of~$\gamma$. Every point in $K$ is a
finite linear combination of curve points, and namely of at most $n+1$
points by Carath\'eodory's theorem. See for instance Ch.~17 of
Rockafellar \cite{\Rocka}.
A consequence of Carath\'eodory's theorem is also that the convex hull
of a compact set in~$\R^n$ is compact; so $K$ is compact. 


We first take care of the case $n\ge3$. 
If the curve is not lying in a hyperplane already, $K$ is an
$n$-dimensional convex body, whose
boundary $\boundary K$ therefore has Hausdorff dimension $n-1\ge2$. It
cannot be filled by a $C^1$-curve, whose image has Hausdorff
dimension~1. Therefore some 
boundary point of $K$ contains a point $P$ that is not on the curve. Let the 
face $F$ be the intersection of $K$ with a supporting hyperplane $\Pi$
at~$P$. 

Now $P$ must be a convex combination of curve points $Q_i$. As
$P\notin\gamma$, this cannot be the trivial convex combination, so 
$P$ cannot be an extreme point of~$K$. Since $F$ is a face, the $Q_i$
(of which there are at least 2) must lie in $F$ as well (and $F$ has
dimension at least 1). 
So we have found at least two curve points $Q_i$ lying in
$F\subset\Pi$. The tangent vectors to~$\gamma$ in the $Q_i$ lie in
$\Pi$ because $\gamma$ lies on a single side of $\Pi$. This proves the
$n\ge3$ part of the lemma. 

Now for the $n=2$ part, $K$ is a 2-dimensional convex compact set,
therefore its boundary $\boundary K$ is a simple closed Lipschitz
curve. If $\boundary K$ has a 1-dimensional face $F$, then $F$ is the
convex hull of two distinct curve points $\x(s_1)$ and $\x(s_2)$, and
the supporting line $\Pi$ through this face is a line of double
tangency. As in the higher dimensional case, this happens in
particular when $\boundary K$ has a point that does not lie on
$\gamma$. 

In the other case, when there is no 1-dimensional face, $K$ is
strictly convex, and every point of $\boundary K\approx S^1$ is a
curve point. If the continuous mapping $S^1\ni s\mapsto
\x(s)\in\boundary K\subset\R^2$ fails to be injective, we again have a
point of double tangency $\x(s_1)=\x(s_2)$ with $s_1\neq
s_2$. However, if the mapping is injective, then it is a
homeomorphism, and $\gamma$ is the boundary of the strictly convex set
$K$. The claim about the monotonic dependence of $s$ on $\theta$
follows routinely from this. 
\qed

{\sc Second Proof of Lemma~\ref{L-curvetouch} (via Borsuk-Ulam):}
Choose $\n\in S^{n-1}$ and maximize
the continuous expression $\x(s)\cdot \n$ over $s\in S^1$ (compact). 
If $s_0$ is the location of a maximum, then the affine hyperplane
$\Pi:=\x(s_0)+ \{\n\}^\perp$ is tangential to $\gamma$ at $\x(s_0)$. 
Now assume the lemma is false. Then for each $\n$,
the maximum of $\x(s)\cdot\n$ is taken on in a unique point. 
In other words, the function $f:S^{n-1}\to
S^1\,,\; \n\mapsto \argmax (\x(s)\cdot \n)$ is well-defined. A routine
argument implies that $f$ is continuous: 
Indeed, let $\n_k\to\n$ and
consider the sequence $(f(\n_k))$. 
If this sequence failed to converge to $f(\n)$, we could
extract a subsequence that stays bounded away from $f(\n)$; but by
compactness it would still have a further subsequence converging to
some quantity $s_*$. Since $\x(f(\n_k))\cdot \n_k \ge \x(s)\cdot \n_k$
for all $s$, we can pass to the limit and conclude
$\x(s_*)\cdot\n\ge \x(s)\cdot\n$ for all $s\in S^1$. But this means
$s_*$ is the (unique) $\argmax\x(s)\cdot\n = f(\n)$, which is a
contradiction. 

Now we focus on $n\ge3$. 
By Borsuk-Ulam, a continuous function from $S^{n-1}$ to $\R^{n-1}$
must map some pair of antipodes into the same point. Using $n\ge3$, we
apply this to the function $f: S^{n-1}\to
S^1\imbeds \R^{n-1}$, obtaining a pair of antipodes $(\n,-\n)$ for which
$f(\n)=f(-\n)=:s_*$. But this means $\max \x(s)\cdot\n
= \min\x(s)\cdot\n =\x(s_*)\cdot\n$,
hence $\x$ lies entirely in the hyperplane $\Pi=\x(s_*)+\{\n\}^\perp$.
This contradiction proves the lemma for $n\ge3$.

Now for $n=2$, we still have the continuous mapping $f:
S^1\ni\n\mapsto\argmax\x(s)\cdot\n\in S^1$. We claim $f$ is
injective. For if it were not, there would exist $s_*$ and two
distinct vectors $\n_1,\n_2$ such that $\x(s)\cdot \n_1\le
\x(s_*)\cdot\n_1$ for all $s$, and likewise $\x(s)\cdot \n_2\le
\x(s_*)\cdot\n_2$. This would make $\x(s_*)$ a curve point in the
vertex of a sector (smaller than a half plane) containing the entire
curve. But this is impossible since $\x(\cdot)$ is $C^1$. 

Now $f:S^1\to S^1$, being continuous and injective, is a homeomorphism,
and $s$ is a monotonic function of the angle of $\n$, or equivalently,
of the angle $\theta$ of the tangent vector, with
$s(\theta+2\pi)=s(\theta)+2\pi$ or $s(\theta+2\pi)=s(\theta)-2\pi$ in
the lift, depending on orientation.
\qed 
 
Note that in the case of curves that are not imbedded but only
immersed in $\R^n$ ($n\ge3$), the second proof guarantees the
existence of `two' tangency points $\x(s_1)$, $\x(s_2)$ with $s_1\neq
s_2$, but does not rule out that this is a double point,
$\x(s_1)=\x(s_2)$; the first proof asserts the slightly stronger
statement $\x(s_1)\neq\x(s_2)$. The weaker version is the one we use. Both
proofs construct a doubly tangent hyperplane $\Pi$ such that the
entire curve lies 
on one side of $\Pi$. This latter property is not of essence for our
purposes. 


{\sc Proof of Thm.~\ref{planar}: } We begin by showing planarity. For
the case of a curve 
$\gamma$ 
with $Z[\psiv]=\emptyset$ (and without loss of generality in~$\R^3$),
we note first that $\gamma$ is real-analytic, as a solution to the EL
equation. We use Lemma~\ref{L-curvetouch} directly to find a plane $\Pi$
tangential to $\gamma$ in two points $\x(s_0)$ and $\x(s_1)$. (We use
only $s_0\neq s_1$, not necessarily $\x(s_0)\neq\x(s_1)$.) The points
$s_1,s_2$ dissect $S^1$ into two open intervals $J_1, J_2$. 
We can now construct another curve
$\tilde{\gamma}:s\mapsto \tilde\x(s)$ by letting $\tilde\x(s)=\x(s)$
for $s\in \bar J_1$ and $\tilde\x(s)=R_\Pi\x(s)$ for $s\in\bar J_2$, where
$R_\Pi$ is the reflection in the plane $\Pi$. The new curve $\tilde\gamma$
is still admissible to (\ref{relOPcurve}); in particular it is still
$C^1$ in $s_{1,2}$, even though its curvature may have jump
discontinuities there. It will carry the same Schr\"odinger
eigenfunction $\phi$. But since $I[\tilde\x,\phi]=I[\x,\phi]$,
$\gamma$ is still a minimizer and therefore $\psiv=\x'\phi$ is a
solution to the EL equations (\ref{EL}). By the unique continuation
property for such solutions (or 
by analyticity in our case), $\gamma=\tilde\gamma$. In other words,
$\gamma|_{J_2}\subset\Pi$. The analogous argument can be made with the
roles of $J_1$ and $J_2$ reversed; so $\gamma\subset\Pi$.

In the case of a D-shaped curve, we take one `corner' point $\x(s_2)$
with $s_2\in Z[\psiv]$ and one regular point $\x(s_1)$
with $s\notin Z[\psiv]$. 
We define the plane $\Pi$ as passing through $\x(s_1)$ and $\x(s_2)$,
and tangential to $\gamma$ at $s_1$. (In case
$\x'(s_1)\parallel \x(s_2)-\x(s_1)$, $\Pi$ is not unique, and any choice
will serve the purpose.) We now define $\tilde\gamma$ as before. The
plane $\Pi$ may intersect $\gamma,\tilde\gamma$ in other points, but 
this is of no concern. $\tilde\gamma$ is still admissible, and is still a
minimizer. The unique continuation argument at $s_1$ guarantees that 
$\gamma$ coincides with $\tilde\gamma$ on the largest interval $J$ that
contains $s_1$ and lies within $Z[\psiv]^c$. So that part of the curve
is planar. But the remaining part is a straight segment connecting the
endpoints of the curve segment $\x(\bar J)$; so the entire curve is
planar. 

The same reflection argument now proves that a planar minimizer cannot
have a double tangent, nor (in the case of a D-shaped minimizer) a
tangent at a regular point $s_1$ that also passes through a point $\x(s_2)$
with $s_2\in Z[\psiv]$. 
According to the lemma, this implies for a regular minimizer (with
$Z[\psiv]=\emptyset$) that $\theta$ is a strictly monotonic function
of~$s$. For a D-shaped minimizer, the first proof of
Lemma~\ref{L-curvetouch} in the case $n=2$ still applies (as $C^1$
wasn't needed), giving that
it is the boundary of a strictly convex set, and that the smooth part
of the curve has the monotonicity property between $s$ and $\theta$.

Finally we want to argue the strengthened statement that the curvature
is actually strictly positive (on the complement of $Z[\psiv]$). To
this end, we use the EL equations 
in polar coordinates, see (\ref{ELpolar}). We have seen that
$s\mapsto\theta(s)$ is strictly monotonic; we can assume
$\theta'(s)\ge0$ without loss of generality (else reflect the
curve). We assume $\theta'(s_*)=0$ for some $s_*$ and try to derive a
contradiction. 

This would make
$s_*$ a minimum of $\theta'$, and therefore $\theta''(s_*)=0$.
From the second of the EL equations, 
$R\theta''+2R'\theta'=\frac{\mu}{R}\sin\theta$, we infer
$\mu\sin\theta(s_*)=0$. 
Now for the system of EL equations (\ref{ELpolar}) with the initial
conditions $\theta(s_*)=\theta_0$ (subject to $\mu\sin\theta_0=0$),
$\theta'(s_*)=0$, $R(s_*)=R_0\neq0$, $R'(s_*)=R_1$, there exists one
solution that can be calculated explicitly, namely
$\theta(s)\equiv\theta_0$, and $R(s)$ solution to the constant
coefficient problem $R''+\lambda R=0$ with the given initial
conditions. By the uniqueness theorem for solutions to regular ODE initial
value problems, this solution is {\em the\/} solution to the EL
equation in question, i.e., $\theta$ is constant, contradicting the
strict monotonicity. 

This proves the theorem. \qed

\section{Asymptotics for Extremals near Singularities}

This section, and the next, are devoted to the proof of
Thm.~\ref{FullRegularity}.  

To avoid trivialities we note first that for a D-shaped minimizer,
$\muv={\bf0}$ is not possible. This is because (\ref{EL}) is trivial
to solve for $\muv={\bf0}$: It gives $\psiv=\left[{
a \cos (\omega s - \alpha) \atop
b \cos (\omega s - \beta) }\right]$ with $\lambda=\omega^2$. 
Such a $\psiv$ can never vanish unless the two components are `in
phase'; but then the range of $\psiv$ would be 1-dimensional.

Our first step is to establish asymptotics near a singularity ($s=0$
with no loss of generality) for solutions to the EL equations. 

\begin{lemma}\label{L-EulerAsy}
Suppose $R,\theta$ satisfy the EL equations \ref{ELpolar} for
$s\in`]0,\ell`[$, with $R>0$ and $\theta'>0$ there, and let $\mu\neq0$, 
Suppose $R(s)\to0$
and $\theta(s)\to\theta_0$ as $s\to0$. Then the following
conclusions hold:

(a)  $\sin\theta_0=0$
\\
(b) $\lim_{s\to0+}R'(s)=:a$ exists, and $a\ge0$. 
   Consequently $\lim_{s\to0}\frac{R(s)}{s}=a$.  
\\
(c) $R(s)\theta'(s)\to0$ and $R^2(s)\theta''(s)\to0$ as $s\to0$.
\end{lemma}

\begin{coro} A D-shaped minimizer would have to be $C^1$.
\end{coro}

{\sc Proof of the lemma: }
We use the assumption that $\lim_{s\to0+}\theta(s)$ exists in the form
that $\theta'$ is integrable on $`]0,\eps]$. 

From the energy estimate, it follows that $R'$ is bounded, and
therefore $R(s)\le bs$ for some $b$. 
Also, $R^2\theta'$ has a finite
limit for $s\to0+$, because it is an antiderivative, on $`]0,\ell`[$,  
of the continuous function $\mu\sin\theta(s)$. If this limit were
non-zero, we would conclude that $\theta'\ge c/s^2$, 
which contradicts the fact that $\theta'$ must be integrable. 

Therefore we know $\lim_{s\to0+} R^2(s)\theta'(s)=0$. 
Again, as an antiderivative of $\mu\sin\theta(s)$, the function 
$R^2\theta'$ is even $C^1$ on $[0,\ell`[$,
and this implies that 
$\lim_{s\to0+} R^2(s)\theta'(s)/s$
exists and is finite. If this limit were nonzero, we would again get
an estimate $\theta'\ge c/s$,  contradicting the integrability of
$\theta'$ near~0.  We have therefore proved 
$
\lim_{s\to0} \frac{R^2(s)\theta'(s)}{s} = 0
$.

Let us pretend to calculate this same limit
in a different manner: the expression is of l'H\^opital type $0/0$,
and $\lim_{s\to0+} \frac{d(R^2\theta')/ds}{ds/ds}$ exists: it is
$\lim_{s\to0+} (R^2\theta')'(s) = \lim_{s\to0+} \mu\sin\theta(s)
= \mu\sin\theta_0$. Combining the two evaluations, we conclude
$\sin\theta_0=0$, proving part (a).

Next, $R\theta'$ is bounded by the energy estimate, and with this, the
Schr\"odinger equation $R''+R(\lambda-\theta'^2)=0$ turns into an
estimate $\abs{R''}\le b+c\theta'$. Therefore $R''$ is integrable up to
$s=0$, and we can extend $R'$ continuously into 0. We have thus proved 
part (b). Trivially $a\ge0$. Below we will see that actually
$a>0$.  

Since we now know that $R'(0)=a$ exists, the energy equation tells us
that $R\theta'$ has a limit as $s\to0+$. If this limit were nonzero,
we would again conclude $\theta'(s)>c/s$ contradicting integrability. 
Now $R^2\theta'' =\mu\sin\theta -2R'\,R\theta'$ will go to 0 as $s\to0$. 
\qed

{\sc Proof of the corollary: }
Part (a) of the lemma implies the corollary. To see this, let us consider the
EL equation (\ref{EL}) over the interval $Z[\psiv]^c$, which we assume
to be $`]0,\ell`[$ with no loss of generality, and test it with 
$\left[{-\psi_2\atop\psi_1}\right]$.  We obtain
$$
\left[\psi_2'\psi_1-\psi_1'\psi_2\right]_{0+}^{\ell-} 
+\int_{0}^{\ell}\muv\cdot\left[{-\psi_2\atop\psi_1}\right]/\abs{\psiv}\,ds
=0
$$
From the boundedness of $\psiv'$ (energy theorem) and the vanishing of
$\psiv$ on the boundary, we obtain that $\muv$ is orthogonal to 
$\int_{0}^{\ell}\left[{-\psi_2\atop\psi_1}\right]/\abs{\psiv}\,ds$,
hence parallel to 
$\int_{0}^{\ell}\psiv/\abs{\psiv}\,ds=\x(\ell)-\x(0)$. On the other
hand, in (\ref{ELpolar}), $\theta$ was the angle between $\psiv$
(hence the curve tangent $\x'$) and
$\muv$ (recall $\muv\neq{\bf0}$). So $\sin\theta_0=0$ means that the
tangent vector $\x'(s)$ becomes parallel to $\x(\ell)-\x(0)$ as
$s\to0$. This could be a $C^1$ curve, or a curve with a cusp (outward
or inward pointing); but a convex curve cannot have cusps. So we have
always $\x'(0)$ in {\em opposite\/} direction as $\x(\ell)-\x(0)$. 
\qed

The cases $\theta_0=0$ and $\theta_0=\pi$ are equivalent under
rotation of the curve by $\pi$, i.e., a shift of $\theta$ by $\pi$ 
(which could instead be absorbed in a
sign change of $\mu$). However, if we have already chosen a preferred direction
of $\muv$ (in the present hypothetical scenario), then the two cases
are distinct: 
\begin{equation}\label{twoDcases}
\raisebox{-47pt}{
\begin{picture}(340,104)
\put(0,0){\includegraphics{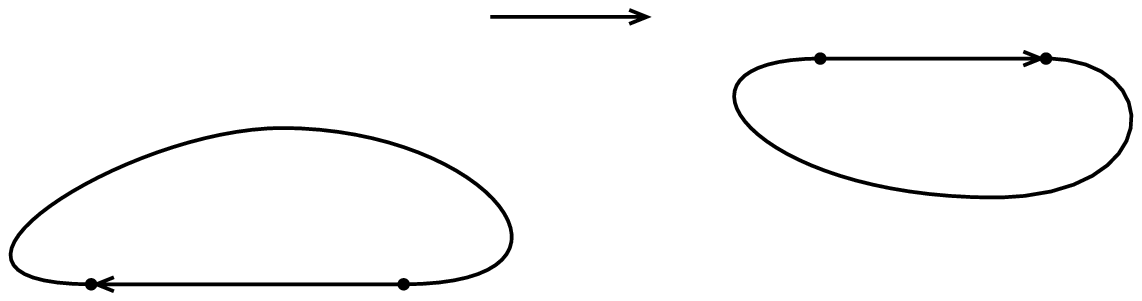}}
\put(110,0){$s=0$}
\put(20,0){$s=\ell$}
\put(162,80){$\muv$}
\put(230,79){$s=0$}
\put(295,79){$s=\ell$}
\put(10,77){$\theta_0=0$}
\put(10,63){$\x'(0)\dirpar\muv$}
\put(230,20){$\theta_0=\pi$}
\put(230, 6){$\x'(0)\antipar\muv$}
\put(45,15){$\x(\ell)-\x(0)$}
\put(245,64){$\x(\ell)-\x(0)$}
\put(170,50){\bf vs.}
\end{picture}
}^{\rule{0pt}{4.5pt}} 
\end{equation}

The key idea for finer asymptotics is that now the angular EL equation 
$R^2\theta''+2RR'\theta'=\mu\sin\theta$ has the same asymptotic
behavior near $s=0$, $\theta=0$ as the Euler equation
$a^2s^2\theta''+2a^2s\theta'=\mu\theta$, 
at least when $a\neq0$. This similarity is brought out by a Sturm
comparison argument, i.e., by integrating the derivative of a mixed
Wronskian, made up of solutions of either equation. 
A variant of this same argument will show that indeed $a>0$. 


\begin{lemma}\label{finethetaasy}
Under the hypotheses of Lemma~\ref{L-EulerAsy}, and $\theta_0=0$, it follows
$a=R'(0)\neq0$. 

Moreover, for $\mu>0$, i.e., the left case in~(\ref{twoDcases}), 
there exist constants $A>0$, $c>0$, 
such that the following finer asymptotics applies for $s\to0+$:
$$
\begin{array}{l}
\theta(s)=   \frac{A}{c} s^c +O (s^{3c}, s^{c+2}) \\
\theta'(s) =  A s^{c-1} + O(s^{3c-1},s^{c+1})\\
R(s) = as + \frac{aA^2}{2c(2c+1)} s^{2c+1} - \frac{\lambda a}{6} s^3  +
o(s^3,s^{2c+1}) 
\\
R'(s) = a + \frac{aA^2}{2c} s^{2c} - \frac{\lambda a}{2} s^2  +
o(s^2,s^{2c}) 
\\
R''(s) = aA^2 s^{2c-1} - \lambda a s  + o(s,s^{2c-1}) 
\end{array}
$$
The case $\mu<0$, i.e., the right case in (\ref{twoDcases}), cannot occur. 
\end{lemma}

{\bf Note: }
For minimizers, it suffices to consider $c\le\frac12$ in
the Lemma. For otherwise, the curvature
$\kappa=\theta'$ would be square integrable, and the Rayleigh quotient
for (\ref{relOPcurve})
could be improved by replacing $R\equiv\phi$ with $\max\{R,\eps\}$,
with the gain of order $\frac{\eps}{a}\times a^2$ coming from $\int
R'^2\,ds$ and a smaller adverse term of order $o(\eps^2)$ coming from
$\int\kappa^2 R^2$.

{\sc Proof of Lemma~\ref{finethetaasy}: }
We first assume $a>0$, $\theta_0=0$, $\mu>0$ and derive the claimed
asymptotics. Thereafter, we will prove that $\mu<0$ leads to a
contradiction. Finally we will lead $a=0$ to a contradiction.

We write the angular EL equation
$\theta''+2\frac{R'}{R}\theta'-\frac{\mu}{R^2}\sin\theta =0$
in the form 
\begin{equation}\label{EL-like-Euler}
s^2 \theta'' + 2 s(1+T_1(s)) \theta' 
   - \Bigl(\frac{\mu}{a^2} + T_2(s)\Bigr)\theta = 0
\;.
\end{equation}
where
$$
T_1(s)= \frac{sR'}{R}-1 = o(1)\;,\quad
T_2(s)= \mu\frac{s^2}{R^2}\,\frac{\sin\theta}{\theta} 
   - \frac{\mu}{a^2}
= O(\theta^2)+O\Bigl(\frac{s^2}{R^2}-\frac{1}{a^2}\Bigr) 
= o(1)
\;.
$$
For comparison, we consider the equation 
$$
s^2 u'' + 2s u' -\frac{\mu}{a^2} u = 0
$$
with its solution $u(s)=s^c$, where $c= -\frac12 +
\sqrt{\frac14+\frac\mu{a^2}}$ is the positive root of the indicial
equation $c(c-1)+2c-\mu/a^2=0$. 
We integrate
\begin{equation}\label{Wronskiprime}
[s^2(\theta' u - \theta u')]' = 
(s^2\theta'' +2s\theta')u - (s^2u''+2su')\theta
\end{equation}
over $[s_1,s_2]\subset`]0,\ell`[$ and obtain
\begin{equation}\label{Wronski}
\left[\rule{0pt}{2.5ex}
s^2(\theta' u - \theta u') \right]_{s_1}^{s_2} = 
\int_{s_1}^{s_2}
\left( T_2\theta - 2sT_1\theta')u\right)\, ds
= \int_{s_1}^{s_2} o(1) (|\theta u| + |s\theta' u|)\,ds
\;.
\end{equation}
We can let $s_1\to0$ and obtain (using that $u,\theta,\theta'>0$)
\begin{equation}\label{Wronskiest}
\abs{s^2(\theta' u - \theta u')} \le \eps
\int_{0}^{s}
 (\theta + s\theta') u \,ds
= \eps s\theta u -\eps \int_0^s s\theta u'\,ds \le \eps s\theta u
\,,
\end{equation}
where $\eps$ can be made as small as we like, provided $s$ is chosen
small. We will want $\eps<c$. 
Dividing, we conclude 
$$
\frac{u'}{u} - \eps\frac1s \le 
\frac{\theta'}{\theta} \le \frac{u'}{u} + \eps\frac1s
$$
on some interval $`]0,\hat s]$.
Integrating again over $[s_1,s_2]\subset `]0,\hat s]$, we get
\begin{equation}\label{ratioest}
\frac{u(s_2)}{u(s_1)} (\frac{s_2}{s_1})^{-\eps} \le
\frac{\theta(s_2)}{\theta(s_1)} 
\le \frac{u(s_2)}{u(s_1)} (\frac{s_2}{s_1})^\eps
\;.
\end{equation}
Fixing $s_2$ and letting $s_1=:s$, this implies for $s<s_2$
$$
\frac{\theta(s)}{s^{c-\eps}}\le \frac{\theta(s_2)}{s_2^{c-\eps}}
\quad\mbox{ and }\quad
\frac{\theta(s)}{s^{c+\eps}}\ge \frac{\theta(s_2)}{s_2^{c+\eps}}
\;.
$$

While this is not yet sufficient to establish the existence of 
$\lim\theta(s)/s^c$, we have at least shown that 
$\theta(s) = O(s^{c-\eps})$. 
Returning to the estimate for $\theta'/\theta$, this implies also
$\theta'(s) = O(s^{c-1-\eps})$.   
This preliminary  estimate will serve to improve the $o(1)$ terms in
$R'=a+o(1)$ and therefore in the Euler equation for $\theta$; redoing
the Sturm comparison with the better estimate will then establish our
desired estimate: 

Namely the Schr\"odinger equation now tells us
$\abs{R''} \le C s^{2c-1-2\eps}$, hence $\abs{R'(s)-R'(0)}\le C
s^{2c-2\eps}$ and $\abs{R(s)-as}\le C s^{2c-2\eps+1}$.   So $\theta$
satisfies (\ref{EL-like-Euler}) with the improved estimates
$$
T_1(s) = O(s^{2c-2\eps})\;,\quad
T_2(s) = O(s^{2c-2\eps})
\;.
$$
This improves our estimates (\ref{Wronski}), (\ref{Wronskiest}),
(\ref{ratioest}) to  
$$
\begin{array}{c}\Dst
\abs{ s^2(\theta'u-\theta u') } 
 \le C \int_0^s (\theta+s\theta')\,u\,s^{2c-2\eps}\, ds 
\le C s\theta u s^{2c-2\eps}
\,,
\\[2ex] \Dst
\frac{u'}{u} - \frac Cs\, s^{2c-2\eps} \le 
\frac{\theta'}{\theta} \le \frac{u'}{u} + \frac Cs\, s^{2c-2\eps}
\,,
\\[2ex]\Dst
\frac{u(s_2)}{u(s_1)} (1-C s_2^{2c-2\eps}) \le
\frac{\theta(s_2)}{\theta(s_1)} \le \frac{u(s_2)}{u(s_1)} (1+C
s_2^{2c-2\eps})
\;.
\end{array}
$$
The constant $C$ does not deteriorate as $s_2\to0$. But for the moment
we fix $s_2$ and conclude (with $s_1=:s$) that 
$$
\limsup_{s\to0} \frac{\theta(s)}{s^c} \le \frac{\theta(s_2)}{s_2^c}
(1-Cs_2^{2c-2\eps})^{-1} 
\quad\mbox{ and }\quad
\liminf_{s\to0} \frac{\theta(s)}{s^c} \ge \frac{\theta(s_2)}{s_2^c}
(1+Cs_2^{2c-2\eps})^{-1} 
\;.
$$
Now we can let $s_2\to0$ and find that
$\lim_{s\to0}\frac{\theta(s)}{s^c}$ exists. We'll call this limit
$A/c$, and it is positive because the  lower
bound for $\liminf\theta(s)/s^c$ was positive. 
Our estimate has also established that $\theta'/\theta\sim c/s$.  
So we have found 
$$
\theta(s)\sim \frac{A}{c} s^c \qquad\mbox{ and }\qquad \theta'(s)\sim A
s^{c-1}
\;.
$$
Inserting these in the Schr\"odinger equation already establishes the
claimed asymptotics for $R''$, and by integration for $R'$, $R$. 

By feeding these asymptotics for $R$ back into the angular EL
equation, we can get a quantitative error term for the asymptotics of
$\theta,\theta'$:
$$
s^2\theta''+2s\theta'-\frac{\mu}{a^2}\theta 
= O(s^{2c+1},s^3)\theta'+O(s^{2c},s^2)\theta = O(s^{3c},s^{2+c})
\;.
$$
Variation of constants quickly establishes $\theta(s) = \frac Ac s^c +
O(s^{3c},s^{c+2})$ and $\theta'(s) = A s^{c-1} + O(s^{3c-1},s^{c+1})$. 

\medskip

In the case $\mu<0$, the indicial equation has no positive roots. We
compare with $u=s^c$ where either
$c=-\frac12+\sqrt{\frac14+\frac{\mu}{a^2}}\in[-\frac12,0`[$ (for
$\frac{\mu}{a^2}\ge-\frac14$), or $c=-\frac12$ (for $\mu<-\frac14$). 
Then  instead of~(\ref{Wronski}), we  obtain
$$
\begin{array}{l}\Dst
\left[\rule{0pt}{2.5ex}
s^2(\theta' u - \theta u') \right]_{s_1}^{s_2} = 
\int_{s_1}^{s_2}
\left( 
\Bigl(\min\{0,{\Tst\frac{\mu}{a^2}-\frac14}\}  + T_2\Bigr)\theta 
- 2sT_1\theta')u\right)\, ds
\\[2ex]\Dst
\phantom{ \left[\rule{0pt}{2.5ex}
  s^2(\theta' u - \theta u') \right]_{s_1}^{s_2} }
\le \int_{s_1}^{s_2} o(1) (|\theta u| + |s\theta' u|)\,ds
\,,
\end{array}
$$
and from this the one-sided estimate 
$$
s^2\theta'u-s^2\theta u'\le \eps s\theta u
\;.
$$
This implies $\frac{\theta'}{\theta}\le \frac{u'}{u}+\frac{\eps}{s}
= \frac{c+\eps}{s}<0$, contradicting the hypothesis $\theta'>0$. 

\medskip

Finally, we consider the case  $a=0$. 
Letting $s\to0$ in the energy theorem, we infer that $E=\mu$, and we
write the energy theorem as
$$
\mu\,(1-\cos\theta)=\frac12(R'^2+R^2\theta'^2+\lambda R^2)
\;.
$$
The case $\mu<0$ immediately forces $R\equiv0$, $\cos\theta\equiv1$, 
since the two sides have opposite signs. 
In the case $\mu>0$ and $a=0$, 
we can proceed similarly as we did for $a>0$, but with an iterative
improvement of the estimate that eventually implies $R(s)\equiv0$ on some
interval $[0,s_*]$, a contradiction. 
We assume $\mu=1$, which is no loss of generality, because can make
$R/\sqrt{\mu}$ into our new function $R$. To begin with, $E=\mu=1$
implies  
$R'^2\le 2(1-\cos\theta)=4\sin^2(\theta/2)$, hence $\abs{R'}\le\theta$.

Let us choose $a_0:=\frac12$ and $s_*$ so that
$\theta(s_*)\le\frac14$ and also
$|R'|\le\frac12a_0$ on $[0,s_*]$. The latter is possible since
$R'\to0$ as $s\to0$. We want to prove inductively 
$R\le\frac12 a_n s$
on $[0,s_*]$ for a sequence $a_n\to0$. 
The start of the induction for $n=0$ follows trivially from integrating
$|R'|\le\frac12a_0$.

Towards an induction step, we use the 
Sturm comparison argument for a one-sided estimate on
$\theta$. Given $a_n\le\frac12$, we let $c_n$ be the positive solution to
$c(c-1)a_n^2+ca_n=\mu=1$, namely
$$
c_n= \frac{-a_n+a_n^2+\sqrt{(a_n-a_n^2)^2+4a_n^2}}{2a_n^2}
\,,
$$
and $u_n:= s^{c_n}$. Note that $a_n\le\frac12$ implies $c_n>1$. 
Then
(dropping the subscript $n$ for a moment)
$$
\begin{array}{l}
[R^2(\theta' u - \theta u')]' 
= 
(R^2\theta''+2RR'\theta')u - (R^2u''+2RR'u')\theta
\\[1.5ex]
\phantom{ [R^2(\theta' u - \theta u')]' }
= 
\sin\theta\,u - \Bigl(\frac{R^2c(c-1)}{s^2} +
\frac{2RR'c}{s}\Bigr)u\theta
\end{array}
$$
Using $\abs{R}\le\frac12a_ns$ and $\abs{R'}\le\frac12a_0s=\frac14s$,
the final parenthesis is $\le\frac14$, so the right hand side is  
$\ge(\frac2\pi-\frac14)\theta u>0$. So we have 
$[R^2(\theta' u - \theta u')]'\ge0$ on $[0,s_*]$.
Integrating from 0 to $s$, we
have $\theta'u-\theta u'\ge0$, hence $\theta'/\theta\ge c/s$ on
$`]0,s_*]$.  Integrating
again from $s$ to $s_*$, we have $\theta(s)\le \theta(s_*)
(\frac{s}{s_*})^c$. 
With the inequality $|R'|\le\theta$ from the energy estimate,
we infer
$$
|R'| \le  \theta(s_*)(\frac{s}{s_*})^c
\quad\mbox{ hence }\quad
|R|\le  \frac{1}{c+1}\, \theta(s_*)(\frac{s}{s_*})^{c+1} s_*
\le  \frac{1}{c}\, \theta(s_*) s \le\frac12\, \frac{1}{2c}s
\;.
$$
Therefore, assuming $R(s)\le\frac12a_n s$ with $a_n\le\frac12$, and
$|R'(s)|\le\frac12a_0=\frac14$, we have concluded
$R(s)\le\frac12a_{n+1}s$, with 
$$
a_{n+1} = 1/(2c_n) =
\frac{a_n^2}{a_n^2-a_n+\sqrt{4a_n^2+(a_n-a_n^2)^2}}
\le \frac{a_n^2}{a_n^2-a_n+2a_n} = \frac{a_n}{a_n+1}
\;.
$$
The positive sequence $(a_n)$ is therefore decreasing and has a limit, 
which has to satisfy $0\le a\le\frac{a}{a+1}$, hence $a=0$. 
Thus $R\equiv0$ on $[0,s_*]$; this contradiction rules out $a=0$.
\qed

\section{Nonexistence of D-shaped Minimizers}
\label{no-D-shaped}

We are now ready to complete the\\ 
{\sc Proof of Theorem~\ref{FullRegularity}:}

Assume we have an extremal $\gamma$ consisting of a strictly
convex curve $(x(s),y(s))$ for $s\in[0,\ell]$, with
$(x(0),y(0))=(0,0)$ and $(x(\ell), y(\ell))= (\ell-2\pi,0)$ and 
$(x'(0),y'(0))=(1,0)=(x'(\ell),y'(\ell))$, and of the straight segment
$[-(2\pi-\ell),0]$ on the $x$-axis. This is understood to include the
case $\ell=2\pi$. 
According to the asymptotics obtained, we may assume that we have
$R(s)\sim as$ for $s\to0+$, with $a>0$, and all the finer results from
Lemma~\ref{finethetaasy}.  As mentioned before, we may and will also
assume $c\le\frac12$, which means that the curvature $\kappa(s)\to\infty$ in
a non-square-integrable manner as $s\to0+$. This simplifies the
mixed-power error terms in Lemma~\ref{finethetaasy}.

We show that such an extremal cannot be a minimizer by giving an
explicit variation that lowers the eigenvalue. The variation we give
is a strong variation, in the sense that, while the change in
$\theta$ is small, the change in $\theta'$ is not. This type of
variation does not enter in the derivation of the EL equation and
provides therefore new information. 

Basically, we connect $\x(0)$ and $\x(\sigma)$, for small $\sigma$'s, with a
comparison curve that preserves the $C^1$ regularity, but whose
curvature stays bounded; and we keep the eigenfunction $R$ constant on
this segment. 
The analogous change is made on the interval
$s\in[-\sigma'+\ell,\ell]$, where we insist that  
$R(\sigma)= R(-\sigma'-\ell)=:R_0$.

Notice that we have the same $a$ at both sides of the segment, 
since $a$ is determined by the energy
theorem $E=\frac12a^2+\mu$. In view of the local asymptotics $R\sim
as$, this ensures that $\sigma'/\sigma\to1$ as $\sigma\to0$. 

The mentioned change in the curve $\x(\cdot)$ may have a slight effect on the
length, which we correct by scaling. 

So we strive to connect the points $(0,0)$ (with horizontal tangent)
to the point $B=(x_0,y_0)$ with slope $\frac{dy}{dx}|_B=m$. We do this
by means of a cubic spline ${\rm Spl}$ given as
$y=k_2x^2+k_3x^3$ with $k_2= (3y_0-mx_0)/x_0^2$, $k_3=(mx_0-2y_0)/x_0^3$.

Specifically in our case, 
$$
\left[\begin{array}{l}x_0\\y_0
\end{array}\right]
=
\gamma(\sigma)
=
\left[
   \begin{array}{l}
      \int_0^{\sigma} \cos\theta(s)\,ds\\ 
      \int_0^{\sigma} \sin\theta(s)\,ds
   \end{array} \right]
=
\left[
   \begin{array}{l}
      \sigma - \frac{A^2}{2c^2(2c+1)}\sigma^{2c+1} +
      O(\sigma^{4c+1}) \\
      \frac{A}{c(c+1)}\sigma^{c+1} 
      +  O(\sigma^{3c+1}) 
   \end{array} \right]
\;,
$$
and
$$
m= \arctan\theta(\sigma) = \frac{A}{c}\sigma^c + O(\sigma^{3c})
\;,
$$
and therefore 
$k_2= \frac{2-c}{c(c+1)}A \sigma^{c-1} + O(\sigma^{3c-1})$ 
and 
$k_3= \frac{c-1}{c(c+1)}A \sigma^{c-2} + O(\sigma^{3c-2})$.
The length of this spline is 
$\int_0^{x_0}\sqrt{1+y'^2}\,dx
=(1+O(\sigma^{2c}))\,(\sigma - O(\sigma^{2c+1}))$, i.e., it differs
from the length $\sigma$ of the original curve 
piece by at most $O(\sigma^{2c+1})$. 

The curvature $\tilde\kappa$ of the spline is 
$$
\tilde\kappa = \frac{y''}{(1+y'^2)^{3/2}} = 
\frac{2k_2+6k_3x}{(1+(2k_2x+3k_3x^2)^2)^{3/2}}
=
(2k_2+6k_3s)\Bigl(1+O(\sigma^{2c})\Bigr) = O(\sigma^{c-1})
\;.
$$
Therefore $\int_{\rm Spl} \tilde\kappa(s)^2\,ds R(\sigma)^2 =
O(\sigma^{2c-1})\sigma^2 = O(\sigma^{2c+1})$.

The new curve $\tilde\gamma$ consists of the cubic spline ${\rm Spl}$
just constructed, the old curve segment
$\gamma|_{[\sigma,\ell-\sigma']}$, an analogous cubic spline ${\rm
  Spl}'$ connecting $(x(\ell-\sigma'),y(\ell-\sigma'))$ to
$(x(\ell),y(\ell))= (\ell-2\pi,0)$, and the straight segment
$[\ell-2\pi,0]$ on the $x$-axis. We consider a new function $\tilde R$
on $\tilde\gamma$ that coincides with $R$ on $[\sigma,\ell-\sigma']$
and is constant $R_0=R(\sigma)=R(\ell-\sigma')$ otherwise. Then
$$
\begin{array}{l}\Dst
\int_{\tilde\gamma} (\tilde R^2\tilde\kappa^2+\tilde R'^2)\,ds -
\int_{\gamma} (R^2\kappa^2+R'^2)\,ds 
\\[1.5ex]\Dst \kern3em\mbox{}\le
\Bigl(\int_{\rm Spl}+\int_{{\rm Spl}'}\Bigr)
    \tilde R^2\tilde\kappa^2\,ds - 
\Bigl(\int_0^\sigma +\int_{\ell-\sigma'}^\ell\Bigr) R'^2\,ds
\\[2.7ex]\Dst \kern3em\mbox{}
\le O(\sigma^{2c+1}) - a^2(\sigma+\sigma') = -2a^2\sigma +
O(\sigma^{2c+1})
\end{array}
$$
Likewise 
$$
\begin{array}{l}\Dst
\int_{\tilde\gamma}\tilde R^2\,ds -\int_{\gamma}R^2\,ds
\ge
(2\pi-\ell)R_0^2 + R_0^2\Bigl(\int_{\rm Spl} + \int_{{\rm Spl}'}\Bigr)ds
- 
\Bigl(\int_0^\sigma+\int_{\ell-\sigma'}^\ell\Bigr) R^2\,ds
\\[2ex]\kern3em\Dst\mbox{}
\ge
(2\pi-\ell)a^2\sigma^2 + a^2\sigma^2(\sigma+\sigma'+O(\sigma^{1+2c})) 
-\frac{a^2}{3}\sigma^3-\frac{a^2}{3}\sigma'^3
\\[1.5ex]\kern3em\Dst\mbox{}
\ge (2\pi-\ell)a^2\sigma^2+ \frac{4a^2}{3}\sigma^3 - O(\sigma^{3+2c})
\ge0
\end{array}
$$
Therefore
$$
\RQ[\tilde\gamma] \le
\RQ[\gamma]-\frac{2a^2\sigma}{\int_\gamma R^2\,ds} + O(\sigma^{2c+1})
\le \lambda - b\sigma
$$
for some $b>0$. Rescaling $\tilde\gamma$ to original length 
introduces a factor $(L[\tilde\gamma]/2\pi)^{2}\le
1+O(\sigma^{2c+1})$, which still leaves us with a competitor whose 
Rayleigh quotient is below $\lambda$. 

This proves that a D-shaped extremal cannot be minimal. 
\qed

\bibliographystyle{plain}
\bibliography{oval}

\end{document}